\documentclass[twocolumn,aps,physrev,showpacs,amsmath,amssymb,superscriptaddress]{revtex4-2}

\usepackage{physics}
\usepackage{graphicx} 
\usepackage[caption=false, labelformat=simple]{subfig}
\usepackage{dcolumn}
\usepackage{csquotes}
\usepackage[colorlinks=true,linkcolor = {blue}, citecolor = {blue}, urlcolor = {blue}]{hyperref}
\usepackage{xcolor}

\usepackage[export]{adjustbox}

\begin{document}

\title{Resonance triplet dynamics in the quenched unitary Bose gas}

\author{J. van de Kraats}
\email[Corresponding author:  ]{j.v.d.kraats@tue.nl}
\affiliation{Eindhoven University of Technology, P. O. Box 513, 5600 MB Eindhoven, The Netherlands}
\author{D. J. M. Ahmed-Braun}
\affiliation{Eindhoven University of Technology, P. O. Box 513, 5600 MB Eindhoven, The Netherlands}
\author{V. E. Colussi}
\affiliation{Infleqtion, Inc., 3030 Sterling Circle, Boulder, CO 80301, USA}
\affiliation{Pitaevskii BEC Center, CNR-INO and Dipartimento di Fisica, Università di Trento, 38123 Trento, Italy}
\author{S. J. J. M. F. Kokkelmans}
\affiliation{Eindhoven University of Technology, P. O. Box 513, 5600 MB Eindhoven, The Netherlands}
\date{\today}

\begin{abstract}
The quenched unitary Bose gas is a paradigmatic example of a strongly interacting out-of-equilibrium quantum system, whose dynamics become difficult to describe theoretically due to the growth of non-Gaussian quantum correlations. We develop a conserving many-body theory capable of capturing these effects, allowing us to model the post-quench dynamics in the previously inaccessible time regime where the gas departs from the universal prethermal stage. Our results show that this departure is driven by the growth of strong lossless three-body correlations, rather than atomic losses, thus framing the heating of the gas in this regime as a fully coherent phenomenon. We uncover the specific few-body scattering processes that affect this heating, and show that the expected connection between the two-body and three-body contacts and the tail of the momentum distribution is obscured following the prethermal stage, explaining the absence of this connection in experiments. Our general framework, which reframes the dynamics of unitary quantum systems in terms of explicit connections to microscopic physics, can be broadly applied to any quantum system containing strong few-body correlations.
\end{abstract}

\maketitle


\textit{Introduction.}--- It is a well known fact that the quantum many-body problem scales exponentially with the number of particles. Hence, despite the rapid rise in available computing power over the last few decades, exact descriptions typically remain out of computational reach \cite{Dirac1929, Yudong2019}. In many cases however, simplifying assumptions can be made about the underlying processes of the system, namely that they obey Gaussian statistics and can be described by long-lived quasiparticles and collective motions of the medium using just a few degrees of freedom \cite{Blazoit1985}. Such simplifications are at the heart of our understanding of a wide range of quantum phenomena, ranging from Fermi liquids and polarons to superfluids and conventional superconductors \cite{Landau1933, Bardeen1957, Anderson1958, Dagotto1994, Fetter2003, Giorgini2008, Pitaevskii2016}.

In realistic quantum systems, however, quasiparticles attain a finite lifetime due to the presence of non-Gaussian correlations, which allow collisions and decay into the many-body continuum \cite{Beliaev1958, Kaplan1976, Pitaevskii1997}. In extreme cases, these correlations lead to such short lifetimes that the quasiparticle spectrum is no longer well defined and the standard simplifications do not apply. The theoretical challenge of understanding such strongly correlated systems is important in many branches of physics, such as condensed matter \cite{Zhitomirsky1999, Hill2001, Dagotto2005, Stone2006, Maniv2015, Cao2018, Meng2023}, ultracold gases \cite{Houcke2012, Schauss2012, Sagi2015,  Chevy2016, Schweigler2017,Shi2018, Ardila2020, Schweigler2021,  Mostaan2023}, nuclear physics \cite{Arrington2012, Hen2017, Berges2021}, and quantum technologies \cite{Dong2008, Gomes2009, Leverrier2011, Bariani2012,  Ra2020, Walschaers2021}.

Due to the large degree of control over the interaction strength, ultracold atomic gases provide a versatile quantum simulator for probing non-Gaussian physics \cite{Haller2009, Chin2010, Bloch2012}. Additionally, the underlying microscopic physics can be encoded exactly in many-body models of these systems, enabling quantitative comparisons between theory and experiment \cite{Kokkelmans2002, Kokkelmans2002_2}. At the frontier of such approaches is the description of ultracold atomic many-body systems featuring non-perturbative few-body effects \cite{Levinsen2015, Fletcher2017, Klauss2017, Christianen2022,  Patel2023, Schumacher2023}. Here, a series of experiments exploring the quench of a degenerate Bose gas to the unitary regime \cite{Makotyn2014, Kira2015_2, Eigen2017, Eigen2018}, where interactions are as strong as allowed by quantum mechanics, raise important fundamental questions regarding the interplay of integrability, ergodicity, and few-body correlations in strongly correlated systems far out of equilibrium. Immediately following the quench, the dynamics are Gaussian and integrable in nature, resulting in the formation of a universal prethermal stage \cite{Regemortel2018, Munoz2019, Gao2020,  Colussi2020}. At later times, non-Gaussian correlations develop and integrability is broken, facilitating the transition of the system towards a global equilibrium. While this behavior is well characterized for weakly interacting systems \cite{Regemortel2018}, the analogous breaking of integrability for quenches to the strongly interacting regime poses a highly nontrivial theoretical problem due to the appearance of non-perturbative phenomena such as strong few-body scattering, bound states of the medium \cite{Musolino2019, Colussi2020, MusolinoPRL}, and rapid three-body losses \cite{Braaten2006}. Furthermore, the quench is associated with the formation of an infinite number of three-body bound Efimov states, whose role and impact remains a subject of active research \cite{Efimov1971, Efimov1979, Colussi2018, Incao2018, MusolinoPRL}.

In this Letter, we utilize the method of cumulants to formulate a conserving theory of the quenched unitary Bose gas, capturing the non-Gaussian correlations forming in the \textit{intermediate time} regime that follows the prethermal stage but preceeds the crossover into a thermal gas. Through direct comparisons with experiment, we characterize the prethermal departure as a consequence of the growth of strong few-body correlations, rather than incoherent particle losses, which broaden the momentum distribution and result in lossless heating of the system. Then, motivated by the absence of a power-law tail in the asymptotics of the momentum distributions observed in Refs.~\cite{Makotyn2014, Eigen2018}, we examine how this expected behavior set by the universal contact relations \cite{Tan2008, Tan2008_2, Tan2008_3, Werner2012, Werner2012_2, Braaten2011, Smith2014} is obscured as the system departs the prethermal regime.


\textit{Model.}---We consider a system of $N$ identical bosons of mass $m$ in a cubic volume $V$, which occupy single-particle states with momentum $\vb{k}$ annihilated by operators $\hat{a}_{\vb{k}}$. Two such atoms may couple to form a energetically-closed channel molecule with center of mass momentum $2\vb{k}$ and internal energy $\nu > 0$, annihilated by the molecular operator $\hat{b}_{2\vb{k}}$. Crucially, we neglect any direct interaction between free atoms, such that all scattering processes in the energetically-open channel are mediated by this molecular state. The resulting Hamiltonian reads \cite{Gurarie2007, Gogolin2008, Schmidt2012},
\begin{eqnarray}
\label{eq:H}
\hat{H} &= \sum_{\vb{k}} \frac{ \hbar^2 k^2}{2m}  \hat{a}_{\vb{k}}^{\dagger} \hat{a}_{\vb{k}} + \sum_{\vb{k}} \left( \frac{ \hbar^2 k^2}{4m}  + \nu\right) \hat{b}_{\vb{k}}^{\dagger} \hat{b}_{\vb{k}} \\ &+ \frac{g}{2\sqrt{V}}\sum_{\vb{k}, \vb{q}} \left[\zeta(2\vb{q}) \  \hat{b}_{\vb{k}}^{\dagger} \hat{a}_{\frac{1}{2} \vb{k}-\vb{q}} \hat{a}_{\frac{1}{2}\vb{k} + \vb{q}}  + \mathrm{h.c.} \right] \nonumber,
\end{eqnarray}
with $k \equiv \abs*{\vb{k}}$. The interaction coupling the atomic and molecular states is modelled by a separable potential with strength $g$ and step-function form factor $\zeta(2 \vb{q}) \equiv \theta(\Lambda - q)$, where $\theta$ is the Heaviside step function such that $\Lambda$ represents a cut-off on the relative two-body momentum. By an analytic solution of the two-body problem, the model parameters can be linked to physical quantities via the renormalization relations \cite{Kokkelmans2002, Braun2022, SupMat},
\begin{equation}
g^2 = \frac{8 \pi \hbar^4}{m^2 R_*}, \qquad \nu = \frac{\hbar^2}{m R_*} \left(\frac{2 \Lambda}{\pi} - a^{-1} \right),
\end{equation}
where $a$ is the $s$-wave scattering length, $R_*$ is the characteristic length scale set by the atom-molecule transition rate \cite{Petrov2004}, and the momentum cut-off is related to the van der Waals length as $\Lambda \sim 1/r_{\mathrm{vdW}}$. At unitarity ($a^{-1} \rightarrow 0$), the dressed molecular state becomes degenerate with the scattering threshold correspondent with a Feshbach resonance \cite{Chin2010}. We express all observables in the Fermi scales $k_n = (6 \pi^2 n)^{1/3}$, $E_n = \hbar^2 k_n^2/2m$ and $t_n = \hbar/E_n$, where $n = N/V$ is the gas density. In a many-body context the relative importance of the molecular state is quantified by the resonance width $R_* k_n$  \cite{Ho2012}.

For both the atomic and molecular fields we adopt the U(1) symmetry breaking picture of a Bose-Einstein condensate (BEC), where the singlets $\expval{a_{\vb{k}}} = \delta_{\vb{k} 0}\sqrt{V} \psi_a$ and $\expval{b_{\vb{k}}} = \delta_{\vb{k} 0}\sqrt{V} \psi_m$ are described by atomic and molecular condensate wave functions $\psi_a$ and $\psi_m$ respectively. To model the quench scenario, we assume that for $t < 0$ the gas is an ideal atomic BEC, such that $\abs{\psi_a}^2 = n$. At $t = 0$, the system is instantaneously quenched to unitarity such that $n a^3 \gg 1$. Then, the far out-of-equilibrium BEC state suffers quantum depletion and sequentially generates higher order correlations in the gas, which we track using a cumulant expansion \cite{Kira2012, Kira2015, Fricke1996, Kohler2002, Kira2014, Kira2015_2}. Truncating the expansion at the level of two-body atomic correlations we obtain the model of Refs.~\cite{Kokkelmans2002_2, Braun2022}, with vanishing background scattering length. This \textit{resonance doublet} model includes the cumulants (denoted with subscript c),
\begin{eqnarray}
n_{\vb{k}}^a &= \expval*{\hat{a}_{\vb{k}}^{\dagger} \hat{a}_{\vb{k}}}_{\mathrm{c}},  \qquad
\kappa_{\vb{k}}^a = \expval*{\hat{a}_{\vb{k}} \hat{a}_{-\vb{k}}}_{\mathrm{c}}, 
\end{eqnarray}
which describe the single-particle momentum distribution and pairing field respectively. Its integrable equations of motion are equivalent to the two-channel Hartree-Fock-Bogoliubov equations \cite{Blazoit1985, Kokkelmans2002_2}. In this work, we introduce an extension of this model referred to as the \textit{resonance triplet} model. Here we include the additional molecular and mixed-channel cumulants,
\begin{eqnarray}
 n_{\vb{k}}^m &= \expval*{\hat{b}_{\vb{k}}^{\dagger} \hat{b}_{\vb{k}}}_{\mathrm{c}}, \qquad \kappa_{\vb{k}}^m = \expval*{\hat{b}_{\vb{k}} \hat{b}_{-\vb{k}}}_{\mathrm{c}},  \\
\chi_{\vb{k}}&= \expval*{\hat{b}_{\vb{k}}^{\dagger} \hat{a}_{\vb{k}}}_{\mathrm{c}}, \qquad\kappa_{\vb{k}}^{am} = \expval*{\hat{b}_{\vb{k}} \hat{a}_{-\vb{k}}}_{\mathrm{c}}, \nonumber
\end{eqnarray}
and the non-Gaussian tripling field,
\begin{equation}
R_{\vb{k},\vb{q}}^a = \expval*{\hat{a}_{\vb{k}} \hat{a}_{\vb{q}} \hat{a}_{-\vb{k} - \vb{q}}}_{\mathrm{c}}.
\end{equation}
In the short-range or vacuum limits, the three-body correlations $\kappa_{\vb{k}}^{am}$ and $R_{\vb{k},\vb{q}}^a$ obey a coupled-channel three-body Schr\"odinger equation and thus represent closed and open-channel components of the three-body wave function \cite{SupMat}. As shown in Ref.~\cite{MusolinoPRL}, $R_{\vb{k},\vb{q}}^a$ becomes macroscopically occupied following the quench and furnishes an order parameter signalling the formation of a Bose-Einstein condensate of Efimovian triples, which motivates its inclusion in the theory. The cumulants $\chi_{\vb{k}}$ and $n_{\vb{k}}^m$ ensure conservation of energy, which is possible in the resonance triplet model due to the cubic form of the Hamiltonian (Eq.~\eqref{eq:H}). This is a crucial difference between the present model and the single-channel triplet model of Ref.~\cite{Colussi2020}, where an inclusion of three-body correlations induces an unphysical leak of energy to the tail of the momentum distribution, thereby prohibiting that model from studying the intermediate time window considered in this Letter.

%
\begin{figure}[t]
\centering
\subfloat{\includegraphics[width=\columnwidth]{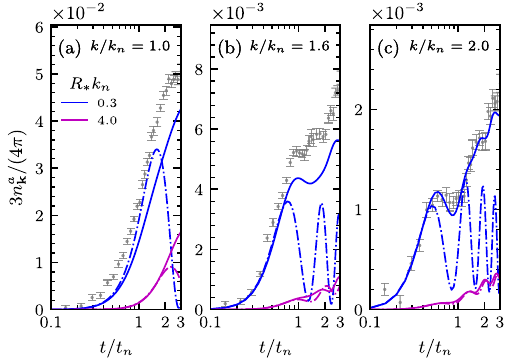}} \newline
\subfloat{\includegraphics[width=\columnwidth]{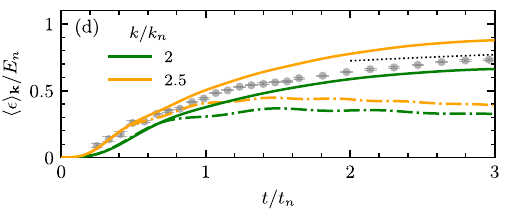}}
\caption{\label{fig:rhok_ek}  Broadening of the single-particle momentum distribution following the quench. In Figs.~(a-c) we plot the dynamics of the excited state population $n_{\vb{k}}^a$ for a set of momenta. Resonance doublet(triplet) model results shown with dash-dotted(solid) lines, matched with the normalization of the experimental data of Ref.~\cite{Eigen2018}, shown with grey circles. In (d) we show in a similar format the dynamics of the average kinetic energy per particle $\expval{\epsilon}_{\vb{k}}$ for the broad resonance case $R_* k_n = 0.3$, comparing with the experimental data of Ref.~\cite{Eigen2017}. Black dotted line shows the expected $T \sim t^{2/13} $ scaling for loss induced heating \cite{Eigen2017}.}
\end{figure}

\begin{figure}[t]
\subfloat{\includegraphics[width=\columnwidth]{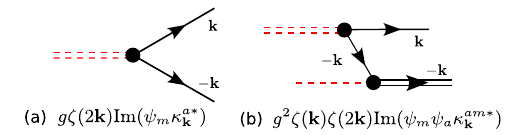}}
\caption{\label{fig:diag} Two-body (a) and three-body (b) scattering processes that drive the depletion of the atomic condensate. Atomic(molecular) states shown with single(double) lines, and condensed states shown in dashed red. Each vertex represents atom-molecule conversion.}
\end{figure}

\textit{Prethermal departure.}--- We begin by examining the dynamics of the momentum distribution in Fig.~\ref{fig:rhok_ek}, comparing against experimental results to look for clues about the causes of the prethermal departure and the nature of the intermediate time regime. Since higher-order correlations develop sequentially following the quench, the doublet and triplet models are initially equivalent \cite{Kira2015}. In this early-time regime, atomic pair excitations are directly generated by condensed molecules, as depicted diagrammatically in Fig.~\ref{fig:diag}(a). As shown in Ref.~\cite{Braun2022}, the early time growth of the condensed molecular fraction scales as $\abs{\psi_m}^2 \sim (t/t_*)^2$, where $t_* = \sqrt{\tau t_n}$ is the mean transition time for atom-molecule conversion. Here $\tau = mR_*/\hbar k_n$ is the molecular lifetime. For broad resonances where $R_* k_n \ll 1$, one finds $t_* \ll t_n$, such that the molecular state acts solely as a mediator of the interaction and the growth of excitations is set purely by $t_n$. If $R_* k_n$ is increased, the associated increase of $t_*$ gradually slows down the excitation of atoms until the narrow resonance limit $R_* k_n\rightarrow \infty$, where all dynamics are frozen.

Following the initial excitation growth, a momentum dependent plateau is reached, signifying the quasi-steady prethermal stage characterized by approximate equilibration of macroscopic observables and emblematic of integrable dynamics \cite{Regemortel2018, Colussi2020}. Our key finding is that the rapid depletion of the condensate for broad resonances spurs the growth of $R^a$ and hence of non-Gaussian three-body correlations, leading to a momentum dependent departure from the prethermal stage visible in Fig.~\ref{fig:rhok_ek}(a-c). The shape of the departure shows remarkable qualitative agreement with the experiment of Ref.~\cite{Eigen2018}, with just slightly slower growth of excitations due to the mismatch in resonance width. In contrast, the integrable resonance doublet model remains in the prethermal stage at all times, consistent with the single-channel doublet model studied in Ref.~\cite{Colussi2020}. 

Physically, the broadening of the momentum distribution shown in Fig.~\ref{fig:rhok_ek} arises from the significant interaction energy injected into the system by the quench, which is gradually converted into kinetic energy via two and three-body scattering and thus results in lossless correlation-induced, rather than recombination induced, heating. To quantify this lossless heating, we follow experiment \cite{Eigen2017}, and examine the averaged kinetic energy per particle \cite{Colussi2020},
\begin{equation}
\expval{\epsilon}_{\vb{k}} = \frac{1}{N} \sum_{  \vb{k}'} n_{\vb{k}'}^a \frac{ \hbar^2 k'^2}{2m} \theta(k - k').
\end{equation}
The restriction to a maximum momentum $k$ accounts for the limited resolution in the experiment \cite{Eigen2017}. From Fig.~\ref{fig:rhok_ek}(d) it is clear that the Gaussian statistics of the resonance doublet model are unable to capture the experimental departure from the prethermal plateau that occurs by $t \sim t_n$.  The resonance triplet model does capture this departure, following the experimental results until the crossover where the gas was experimentally found to pass from degenerate to nondegenerate regimes. This agreement further strengthens the interpretation of the dynamics at intermediate times as correlation (rather than loss) dominated.

\begin{figure}[t]
\includegraphics[width=\columnwidth]{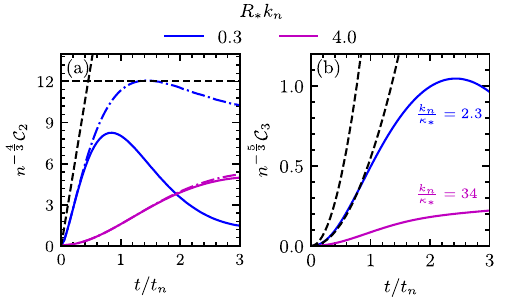}
\caption{\label{fig:CondFrac}  Dynamics of two-body and three-body contact densities $\mathcal{C}_2$ in (a) and $\mathcal{C}_3$ in (b), for different values of the resonance width $R_* k_n$. Results from the resonance doublet(triplet) model shown with dash-dotted(solid) lines. For comparison we show with black dashed lines the linear early time growth of $\mathcal{C}_2$ as derived in Ref.~\cite{Corson2015} for broad resonances and the long time constant value of $\mathcal{C}_2$ from Ref.~\cite{Sykes2014}. For $\mathcal{C}_3$ we show the range of quadratic early time growths derived in Ref.~\cite{Colussi2018}, which correspond to maximum ($k_n/\kappa_* \sim 1$) and minimum ($\kappa_* \ll k_n$) enhancement  of $\mathcal{C}_3$ due to the Efimov effect, again in the broad resonance limit.}
\end{figure}

\textit{Correlations in the intermediate time regime.}--- Having provided evidence for the lossless nature of the intermediate time regime, we now study the dominant processes in the system during this time. The difference between the resonance doublet and resonance triplet dynamics in Fig.~\ref{fig:rhok_ek} arises predominantly from a distinct non-Gaussian process in the resonance triplet model, shown diagrammatically in Fig.~\ref{fig:diag}(b) \cite{SupMat}. Together, the processes in Fig.~\ref{fig:diag} seed the post-quench growth of short-range atomic two- and three-body correlation functions, given as,
\begin{equation}
\expval{d^{\dagger} d} = \abs{\frac{1}{V} \sum_{\vb{k}} \kappa_{\vb{k}}^{a}}^2, \qquad \expval{t^{\dagger} t} = \abs{\frac{1}{V^{\frac{3}{2}}} \sum_{\vb{k}, \vb{q}} R_{\vb{k},\vb{q}}^a}^2.
\end{equation}
Here $\hat{d} = \hat{\psi}(\vb{0})\hat{\psi}(\vb{0})$ and $\hat{t} = \hat{\psi}(\vb{0})\hat{\psi}(\vb{0}) \hat{\psi}(\vb{0})$ represent the local dimer and trimer field respectively, with the field operator $\hat{\psi}(\vb{0}) = (1/\sqrt{V}) \sum_{\vb{k}} \hat{a}_{\vb{k}}$. Due to their short-range nature, the value of the correlation functions is expected to be fully specified by universal relations derived from few-body physics. Hence they can be directly related to the two-body and three-body contact densities $\mathcal{C}_2$ and $\mathcal{C}_3$, which form a measure of the probability for finding two and three particles in close proximity \cite{Tan2008, Tan2008_2, Tan2008_3, Braaten2011}. In our two-channel model, we derive the following relations from effective field theory \cite{SupMat},
\begin{eqnarray}
&\mathcal{C}_2 &= \frac{m^2 g^4}{4 \hbar^4 \nu^2} \expval{d^{\dagger} d} +\frac{m^3 g^6}{2 \hbar^6 \nu^3 \Lambda^2} \left(H + \frac{J}{\pi} \right) \expval{t^{\dagger} t} \label{eq:contacts1}\\
&\mathcal{C}_3 &= -\frac{m^2 g^4}{8 \hbar^4 \nu^2 \Lambda^2} H' \expval{t^{\dagger} t}.
\label{eq:contacts2}
\end{eqnarray}
Here $H, J, H'$ are known log-periodic functions of the Efimovian binding wavenumber $\kappa_*$ and the cut-off $\Lambda$ \cite{Braaten2011}. As shown in Fig.~\ref{fig:CondFrac}, the prethermal departure is indeed correlated with a significant increase in $\mathcal{C}_3$, indicating the introduction of strong non-Gaussian three-body correlations in this regime. Simultaneously, the influence of surrounding particles on clustered pairs leads to a significant decrease of the two-body contact \cite{Colussi2020}. As shown in Ref.~\cite{MusolinoPRL}, the value of $\mathcal{C}_2$ and $\mathcal{C}_3$ following the quench is determined predominantly by the magnitude of macroscopic order parameters associated with condensation of pairs and triples. These findings are confirmed in our model by the development of a substantial fraction of condensed triples for $t > t_n$ \cite{SupMat}. We note that the increase of $\mathcal{C}_3$ for broad resonances is connected with an increased overlap of the size of the Efimov trimer, quantified by the binding wavenumber $\kappa_*$, and the Fermi scale $k_n$. Consistent with Figs.~\ref{fig:rhok_ek}(a-c), the value of $\mathcal{C}_3$ is decreased for narrow resonances, where $\kappa_* \ll k_n$ represents the previously unexplored opposite limit to the universal regime $\kappa_* \gg k_n$ examined in Ref.~\cite{MusolinoPRL}.

\begin{figure}[t]
\includegraphics[width=\columnwidth]{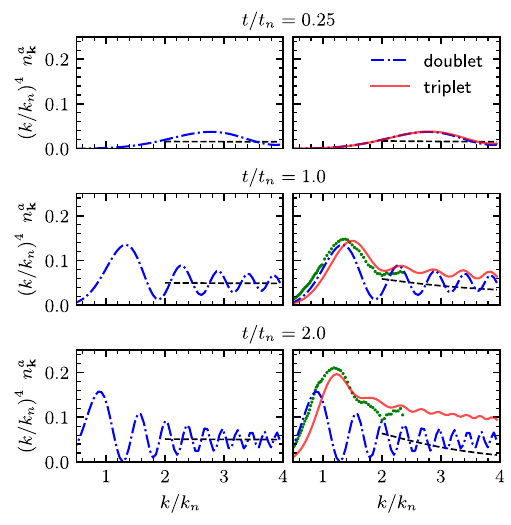}
\caption{\label{fig:tails} Dynamics of the tail of the single-particle momentum distribution for the broad resonance $R_* k_n = 0.3$. In the left(right) panels we compare the resonance doublet(triplet) models with the asymptotic prediction in Eq.~\eqref{eq:tail}, shown with a black dashed line. For the sake of comparison the doublet results are replotted in the right-hand panels. For $t/t_n = 1.0$ and $2.0$ we also compare with the experimental data of Ref.~\cite{Eigen2018}, shown with the green circles.}
\end{figure}

%
%

\textit{Asymptotics in the intermediate time regime.}---We now focus on the large-momentum tail of $n_{\vb{k}}^a$, whose expected behavior due to universal relations derived from few-body physics was not observed in experiments \cite{Makotyn2014, Eigen2018}. Specifically, the following asymptotic behavior is predicted at thermal equilibrium,
\begin{equation}
k^4 n_{\vb{k}}^a \underset{k \rightarrow \infty}{\longrightarrow} \mathcal{C}_2 + \mathcal{C}_3 F(k)/k,
\label{eq:tail}
\end{equation}
where $F(k)$ is a log-periodic function specified by $\kappa_*$ \cite{Braaten2011}. While Eq.~\eqref{eq:tail} provides a possible route to extract the values of the contacts from $n_{\vb{k}}^a$, a $1/k^4$ tail has so far not been observed in quench experiments \cite{Makotyn2014, Eigen2018}. Fits to Eq.~\eqref{eq:tail} were made in Ref.~\cite{Smith2014}, but found values for $\mathcal{C}_2$ considerably larger than expected from theoretical calculations \cite{Diederix2011, Sykes2014, Colussi2020}. Additionally it has been shown that in lower dimensional systems the scaling in Eq.~\eqref{eq:tail} can be disrupted by non-local correlations \cite{Corson2016}.

Motivated by this disagreement between existing theory and experiment, we compare in Fig.~\ref{fig:tails} the single-particle momentum distribution obtained from our models with the asymptotic prediction in Eq.~\eqref{eq:tail}. In the resonance doublet model, where all dynamics are Gaussian, $\mathcal{C}_3$ vanishes trivially and one can show analytically that the expected $1/k^4$ power law due to $\mathcal{C}_2$ is always present \cite{SupMat}. In the resonance triplet model, we observe that the non-Gaussian processes responsible for the departure from the prethermal stage for $t \gtrsim t_n$ damp the oscillatory behavior of $n_{\vb{k}}^a$, consistent with the findings of Ref.~\cite{Regemortel2018} at weak interactions and in agreement with the experimental data of Ref.~\cite{Eigen2018}. At the same time, the significant increase of excitations for $k > k_n$ obscures the expected power laws for all the examined momenta $k/k_n \leq 4$. Hence, once the gas has exited the prethermal stage, the asymptotic expansion in Eq.~\eqref{eq:tail} no longer captures the momentum distribution over the considered range. As experimental results are lacking beyond this range due to poor signal-to-noise ratio \cite{Makotyn2014}, our findings suggest that significant caution should be exercised when fitting Eq.~\eqref{eq:tail} out of equilibrium, and explain why a power-law tail was not seen in Refs.~\cite{Makotyn2014, Smith2014}. It is important to note however that our results do not invalidate Eq.~\eqref{eq:tail}, but rather push its possible applicability to larger momenta. For sufficiently large values of $k/k_n$, our distributions do converge consistently with the value of $\mathcal{C}_2$ obtained from Eq.~\eqref{eq:contacts1}, but also exhibit strong finite range features due to the density regimes considered \cite{SupMat}. We have confirmed that the distributions in Fig.~\ref{fig:tails} are density independent.
%
%

\textit{Conclusion}--- In this Letter, we have shown how a conserving many-body model constructed from a selection of microscopically relevant Gaussian and non-Gaussian correlations, is able to elucidate the dynamics of quenched unitary Bose gases in the time window succeeding the integrable dynamics associated with the prethermal stage. In the future, this general framework can be used in studying the wide-array of other quantum systems that exhibit non-Gaussian physics, including for example strongly interacting ultracold mixtures and polarons \cite{Patel2023, Schumacher2023, Hu2016, Duda2023, Mostaan2023}, trions in semiconductors \cite{Mak2013, Deilmann2016, Rana2020, Rana2021, Zhumagulov2022}, nuclear matter \cite{Bethe1965, Dean2003, Otsuka2020, Perez2023}, and Rydberg atom arrays \cite{Deger2023}.

\begin{acknowledgments}
We thank Silvia Musolino for fruitful discussions. J.v.d.K. and S.J.J.M.F.K. acknowledge financial support from the Dutch Ministry of Economic Affairs and Climate Policy (EZK), as part of the Quantum Delta NL program. D.J.M.A.B. acknowledges financial support from the Netherlands Organisation for Scientific Research (NWO) under Grant No. 680-47-623. V.E.C. acknowledges financial support from the Provincia Autonoma di Trento, the Italian MIUR under the PRIN2017 projectCEnTraL, and from the National Science Foundation (NSF) under Grant No. NSF PHY-1748958.
\end{acknowledgments}

\bibliography{References}

\end{document}


\title{Supplementary Material: ``Resonance triplet dynamics in the quenched unitary Bose gas"}

\author{J. van de Kraats}
\email{j.v.d.kraats@tue.nl}
\affiliation{Eindhoven University of Technology, P. O. Box 513, 5600 MB Eindhoven, The Netherlands}
\author{D.J.M. Ahmed-Braun}
\affiliation{Eindhoven University of Technology, P. O. Box 513, 5600 MB Eindhoven, The Netherlands}
\author{V.E. Colussi}
\affiliation{Infleqtion, Inc., 3030 Sterling Circle, Boulder, CO 80301, USA}
\affiliation{Pitaevskii BEC Center, CNR-INO and Dipartimento di Fisica, Università di Trento, 38123 Trento, Italy}
\author{S.J.J.M.F. Kokkelmans}
\affiliation{Eindhoven University of Technology, P. O. Box 513, 5600 MB Eindhoven, The Netherlands}
\date{\today}

\begin{abstract}
\end{abstract}
\maketitle

\section{TWO-BODY INTERACTION}
\label{sec:renorm}

In this section we provide further detail regarding the two-body interaction used in the Hamiltonian in Eq.~(1) in the main text, see also Ref.~\cite{Braun2022}. We consider the stationary two-body scattering problem in the basis of states $\ket*{\vb{k}, -\vb{k}}$ describing two counter propagating plane waves with relative momentum $\vb{k}$. In addition we define an energetically open and energetically closed channel, correspondent with distinct internal states of the particles, and fix the internal energy of the open channel to zero. We adopt the Feshbach formalism \cite{feshbach1992}, defining projection operators $\hat{\mathcal{P}}$ and $\hat{\mathcal{Q}}$ onto the open channel and closed channel subspaces respectively, with properties $\hat{\mathcal{P}}\hat{\mathcal{P}} = \hat{\mathcal{P}}, \hat{\mathcal{Q}}\hat{\mathcal{Q}} = \hat{\mathcal{Q}}, \hat{\mathcal{P}}+\hat{\mathcal{Q}} = \mathbb{I}$ and $\hat{\mathcal{P}}\hat{\mathcal{Q}} = 0$. A general two-body operator $\hat{O}$ may be decomposed as,
\begin{align}
\begin{split}
\hat{O} = \hat{\mathcal{P}}\hat{O} \hat{\mathcal{P}} + \hat{\mathcal{P}}\hat{O} \hat{\mathcal{Q}} + \hat{\mathcal{Q}}\hat{O} \hat{\mathcal{P}} + \hat{\mathcal{Q}}\hat{O} \hat{\mathcal{Q}} \equiv \hat{O}_{PP} + \hat{O}_{PQ} + \hat{O}_{QP} + \hat{O}_{QQ} 
\end{split}
\end{align}
We consider the atomic projection $\hat{\mathcal{P}} \ket*{\psi} = \ket*{\psi_P}$, where $\hat{H} \ket*{\psi} = E \ket*{\psi}$. By inserting projection operators one can rewrite to obtain,
\begin{align}
\begin{split}
\left(\hat{H}_{PP} + \hat{V}_{PQ} \hat{G}_{Q}(E) V_{QP}\right) \ket*{\psi_P} = E \ket*{\psi_P},
\end{split}
\label{eq:SchrodingerProjected}
\end{align}
where $\hat{V}_{PQ} = \hat{V}_{QP}$ is the two-body interaction coupling the open and closed channel subspaces, and $\hat{G}_Q(E) = (E - \hat{H}_{QQ})^{-1}$ the closed-channel Green's function. Equation \eqref{eq:SchrodingerProjected} motivates the introduction of an \textit{effective} potential,
\begin{align}
\begin{split}
\hat{V}_{\mathrm{eff}} =  \hat{V}_{PQ} \hat{G}_{Q}(E) V_{QP} = \hat{V}_{PQ} \frac{\ket*{\phi} \bra*{\phi}}{E - \nu} V_{QP}.
\label{eq:Veff}
\end{split}
\end{align}
Here we have made the important assumption that the resonant bound state $\ket*{\phi}$ is the sole eigenstate of the closed-channel Hamiltonian, such that $\hat{G}_Q(E) = (E - \nu)^{-1} \ket*{\phi} \bra*{\phi}$. This isolated resonance approximation is valid provided that the state $\ket*{\phi}$ is well separated from other closed-channel states and its binding energy is large compared to the scattering energy. From Eq.~\ref{eq:Veff} we formulate the open-channel component of the two-body transition matrix,
\begin{align}
\begin{split}
\matrixel{\vb{k}', -\vb{k}'}{\hat{t}_{PP}(E+i0)}{\vb{k}, -\vb{k}} =  \matrixel{\vb{k}', -\vb{k}'}{\hat{V}_{PQ}\frac{\ket*{\phi} \bra*{\phi}}{E + i0 - \nu} V_{QP}}{\psi_P}
\label{eq:ktk}
\end{split}
\end{align}
which we evaluate for infinitesimally complex energies to skirt the poles on the real axis, and where $E = \hbar^2 \abs{\vb{k}'}^2/m = \hbar^2 \abs{\vb{k}}^2/m$. As noted in the main text, we model the potential by a single term separable form $\hat{V}_{PQ} = \hat{V}_{QP} = \frac{1}{\sqrt{V}} \beta \ket*{\zeta} \bra*{\zeta}$, where the form factors are given as $\braket*{\vb{k}, -\vb{k}}{\zeta} = \zeta(2\vb{k}) = \theta(\Lambda - \abs{\vb{k}})$. With the ansatz $\hat{t}_{PP}(z) = \ket*{\zeta} \tau_{PP}(z) \bra*{\zeta}$, Eq. \eqref{eq:ktk} can be formally solved to obtain, 
\begin{align}
\begin{split}
\tau_{PP}(E+i0) =\frac{g^2}{2V} \frac{1}{E +i0 - \nu - \frac{g^2}{2V}  \matrixel*{\zeta}{\hat{G}_0^{\mathrm{2B}}(E+i0)}{\zeta}},
\end{split}
\label{eq:tau_open}
\end{align}
where we have defined $g = \sqrt{2} \beta \braket*{\phi}{\zeta}$ as the coupling strength. The matrix elements of the uncoupled two-body Green's function $\hat{G}_0^{\mathrm{2B}}(z) = (z - \hat{H}_0)^{-1}$ are evaluated as,
\begin{align}
\begin{split}
\matrixel*{\zeta}{\hat{G}_0^{\mathrm{2B}}(E + i0)}{\zeta} = \frac{m V}{2\pi^2 \hbar^2} \left[ - \Lambda + k \ \mathrm{arctanh}\left( \frac{\Lambda}{k}\right) \right].
\end{split}
\end{align}
From Eq. \eqref{eq:tau_open}, we can derive the low-energy behavior of the open-channel scattering amplitude,
\begin{align}
\begin{split}
f(E) = -\left[ \frac{1}{a} + ik + \frac{1}{2} \left(2 R_* - \frac{4}{\pi \Lambda} \right)k^2 \right]^{-1},
\end{split}
\end{align}
which holds to second order in $k/\Lambda$. Here we have introduced the scattering length $a$ and resonance length $R_*$ as,
\begin{align}
\begin{split}
\frac{1}{a} = \frac{2 \Lambda}{\pi} - \frac{8 \pi \hbar^2 \nu}{m g^2}, \qquad R_* = \frac{8 \pi \hbar^4}{m^2  g^2},
\label{eq:aRs}
\end{split}
\end{align}
which may be inverted to obtain the renormalization relations quoted in Eq.~(2) in the main text. The above expression for $R_*$ follows the definition in Ref. \cite{Petrov2004}. Finally we can relate $\Lambda$ to the range of the potential by considering the energy $E_D$ of the Feshbach dimer, signified by a pole in the transition matrix for $a > 0$. If we define a dimer wavenumber $\kappa_D = \sqrt{m \abs{E_D}}$, then a low-energy expansion of $E_D$ reveals that,
\begin{align}
\begin{split}
\Lambda = \frac{2}{\pi} \frac{\kappa_D^2}{\kappa_D + R_* \kappa_D^2 - \frac{1}{a}},
\end{split}
\label{eq:LambdaKappa}
\end{align}
to second order in $\kappa_D/\Lambda$. In systems with van der Waals interactions, the dimer binding energy near resonance is universally determined as $\kappa_D^2 = (a - \bar{a} + R_*)^{-2}$ \cite{Chin2010}, where $\bar{a} = 0.995978 \ r_{\mathrm{vdW}}$ is the mean scattering length \cite{Gribakin1993}. Substituting into Eq. \eqref{eq:LambdaKappa} one finds,
\begin{align}
\begin{split}
\Lambda = \frac{2}{\pi} \frac{1}{\bar{a} \left(1 - \frac{\bar{a}}{a} + 2 \frac{R_*}{a} - \frac{R_*^2}{\bar{a} a} \right)} \approx \frac{2}{\pi \bar{a}},
\end{split}
\end{align}
where the last equality holds in the resonant limit $a \gg \bar{a}, R_*$, which is always true in our calculations. The universal zero-range limit is achieved when $\Lambda$ far exceeds all other momentum scales, which has the important implication in our model that the ratio $r_{\mathrm{vdW}}/R_*$ should be small to avoid non-universal finite range effects. For all computations in the main text we set $\Lambda$ such that $n r_{\mathrm{vdW}}^3 = 5.8 \cdot 10^{-7}$, which is sufficiently dilute for our results to be universal in the regime of resonance strengths we examine.

\section{Many-body model}
\label{sec:eom}
In this section we provide further detail regarding our many-body formalism and the cumulant equations of motion.

\subsection{Molecular operators}

If we define $\hat{c}_{\vb{k}}$ as the operator that annihilates a particle with momentum $\vb{k}$ in the closed spin-channel, then the composite molecular operator is defined as \cite{Altman2005},
\begin{align}
\begin{split}
\hat{b}_{\vb{k}} = \sum_{\vb{q}} \frac{\phi(\vb{q})}{\sqrt{2}}  \hat{c}_{\frac{\vb{k}}{2} + \vb{q}} \hat{c}_{\frac{\vb{k}}{2} - \vb{q}},
\end{split}
\label{eq:b_def}
\end{align}
where $\phi(\vb{q}) \equiv \braket*{\vb{q}}{\phi} $. The operator $\hat{b}_{\vb{k}}$ obeys canonical commutation relations provided that the state $\ket*{\phi}$ is strongly localized with respect to the density scales of the gas \cite{Braun2022}. Since we assume that the state $\ket*{\phi}$ is the only eigenstate of the closed channel Hamiltonian $H_{QQ}$, Eq.~
\ref{eq:b_def} may be inverted as $\hat{c}_{\frac{\vb{k}}{2} + \vb{q}} \hat{c}_{\frac{\vb{k}}{2} - \vb{q}} = \sqrt{2} \phi(\vb{q}) \hat{b}_{\vb{k}}$. Hence we can convert any expectation value containing an even-numbered product of single-particle closed-channel operators into a product of molecular operators, whilst odd-numbered products vanish.

\subsection{Cumulant equations of motion}

To track the post quench dynamics, we solve the Heisenberg equations of motion ,
\begin{align}
\begin{split}
i \hbar \pdv{t} \expval*{\hat{O}_p} = \expval*{\left[\hat{O}_p, \hat{H} \right]},
\end{split}
\label{eq:HB}
\end{align}
with $\hat{O}_p$ an arbitrary $p$-body operator and $\hat{H}$ given by Eq. (1) of the main text. To separate out genuine few-body correlations from expectation values, we introduce the cumulant of $\hat{O}_p$ as \cite{Fricke1996, Kohler2002},
\begin{align}
\expval{\hat{O}_p}_c = \expval{\prod_{i=0}^l \hat{\alpha}_{\vb{k}_i}^{\dagger} \prod_{j=0}^m\hat{\beta}_{\vb{k}_j'}}_c = (-1)^m \prod_{i=0}^l \pdv{x_i} \prod_{j=0}^m \pdv{y_j^*} \ln \expval{e^{\sum_{i=0}^l x_i \hat{\alpha}_{\vb{k}_i}^{\dagger}} e^{\sum_{j=0}^m y_j^* \hat{\beta}_{\vb{k}_j'}}} \Bigg|_{\vb{x}, \vb{y} = 0},
\label{eq:cudef}
\end{align}
where $\hat{\alpha}, \hat{\beta}$ can represent either atomic or molecular operators. The resulting cumulant expansion transforms Eq. \eqref{eq:HB} to a system of coupled ordinary differential equations for all the individual cumulants. Due to the cubic structure of $\hat{H}$, a cumulant containing $n$ operators couples to cumulants containing at most $n+1$ operators. This is a major simplification compared to the more commonly used quartic single-channel Hamiltonian, where coupling occurs also to cumulants with $n+2$ operators \cite{Colussi2020}.

The dynamics of the atomic and molecular condensate wave functions are given by the two-channel Gross-Pitaevskii equations,
\begin{align}
i \hbar \pdv{t}\psi_a &= g \left(\zeta(0) \psi_m \psi_a^* + \frac{1}{V}\sum_{\vb{q}}  \zeta(\vb{q}) \chi_{\vb{q}}^* \right), \label{eq:psia} \\
i \hbar  \pdv{t}\psi_m &= \nu \psi_m +  \frac{g}{2} \left(\zeta(0) \psi_a^2 + \frac{1}{V}\sum_{\vb{q}} \zeta(2\vb{q}) \kappa_{\vb{q}}^a \right). \label{eq:psim}
\end{align}
Upon including the atomic excitation density and atomic pairing field as defined in Eq.~(3) in the main text, one obtains the two-channel Hartree-Fock-Bogoliubov equations,
\begin{align}
i \hbar \pdv{t}n_{\vb{k}}^a &= g \zeta(2\vb{k}) \left(\psi_m \kappa_{\vb{k}}^{a*} - \psi_m^* \kappa_{\vb{k}}^{a} \right)  - g \zeta(\vb{k}) \left(\psi_a \chi_{\vb{k}} -\psi_a^* \chi_{\vb{k}}^*\right), \label{eq:nka}\\
i \hbar  \pdv{t}\kappa_{\vb{k}}^a &= 2 \varepsilon_{\vb{k}}^a \kappa_{\vb{k}}^a + g\zeta(2\vb{k}) \left[ (1 +  n_{\vb{k}}^a )  (1 +  n_{\vb{k}}^a ) \psi_m -\left(n_{\vb{k}}^{a}\right)^2 \psi_m \right] +  2g \zeta(\vb{k}) \kappa_{\vb{k}}^{am}  \psi_a^* \label{eq:kappaa}.
\end{align}
Here $\varepsilon_{\vb{k}}^a = \hbar^2 k^2/(2m)$ gives the atomic kinetic energy. In the equation of motion for $\kappa_{\vb{k}}^a$, we recognize the Bose enhancement factors $(1+n_{\vb{k}}^a)$, which alter the coupling to excited states with non-zero occupation. In the resonance triplet model, the additional cumulants introduced in Eqs.~(4) and (5) in the main text obey the equations of motion,
\begin{align}
i \hbar  \pdv{t}\kappa_{\vb{k}}^{am} &= \left(\varepsilon_{\vb{k}}^a + \varepsilon_{\vb{k}}^m \right)\kappa_{\vb{k}}^{am} +g \zeta(\vb{k})\left( \kappa_{\vb{k}}^a \psi_a +  \kappa_{\vb{k}}^m\psi_a^* \right) + g \zeta(2\vb{k}) \chi_{\vb{k}}^* \psi_m  + \frac{g}{2\sqrt{V}} \sum_{\vb{q}} \zeta( 2\vb{q} + \vb{k}) R_{\vb{k},\vb{q}}^a,  \\
i \hbar  \pdv{t} \chi_{\vb{k}} &= \left(\varepsilon_{\vb{k}}^a - \varepsilon_{\vb{k}}^m \right) \chi_{\vb{k}} - g \zeta(\vb{k})  \left(n_{\vb{k}}^a - n_{\vb{k}}^m \right) \psi_a^* + g\zeta(2\vb{k}) \kappa_{\vb{k}}^{am*} \psi_m, \label{eq:chi} \\
i \hbar \pdv{t}n_{\vb{k}}^m &= g \zeta(\vb{k}) \left(\psi_a \chi_{\vb{k}} -\psi_a^* \chi_{\vb{k}}^*\right), \\
i \hbar  \pdv{t}\kappa_{\vb{k}}^m &= 2 \varepsilon_{\vb{k}}^m  \kappa_{\vb{k}}^m + 2 g \zeta(\vb{k}) \kappa_{\vb{k}}^{am} \psi_a, \\
i  \hbar \pdv{t}R_{\vb{k},\vb{q}}^a &= \mathcal{\hat{S}}\bigg\{ \varepsilon_{\vb{k}}^a  R_{\vb{k},\vb{q}}^a + \frac{g}{\sqrt{V}}(1 + n_{\vb{k}}^a + n_{\vb{q}}^a) \zeta(\vb{k} - \vb{q}) \kappa_{\vb{k} + \vb{q}}^{am} \bigg\}
\end{align}
Here $\mathcal{\hat{S}}$ is a three-body symmetrization operator defined as $\mathcal{\hat{S}} = 1 + \hat{P}_+ + \hat{P}_-$, where $\hat{P}_{+(-)}$ define (anti)symmetric cyclic permutation operators of the three particle indices. We have also defined the molecular kinetic energy $\varepsilon_{\vb{k}}^m = \hbar^2 k^2/(4m) + \nu$.

The molecular and anomalous excitation densities $n_{\vb{k}}^m$ and $\chi_{\vb{k}}$ are necessary inclusions in the model to ensure that the total system energy is always conserved. If these cumulants are excluded, the growth of the three-body cumulant $\kappa_{\vb{k}}^{am}$ leads to an unphysical increase in the total kinetic energy, similar to the violation observed in single-channel triplet models \cite{Colussi2020}. As we show in Sec.~\ref{sec:excgrowth}, the cumulants $n_{\vb{k}}^m$ and $\chi_{\vb{k}}$ are actually fully determined from short-range physics, meaning that they have predominant amplitude in the large momentum modes $k \sim \Lambda$. This necessarily implies that the value of these cumulants will be strongly sensitive to the short-range detail of the two-body interaction, and thus non-universal in nature. For this reason we do not assign physical significance to the values of $n_{\vb{k}}^m$ and $\chi_{\vb{k}}$, but treat them as necessary quantities required to regularize the short-range behavior of the theory. A detailed analysis of the dynamics of $n_{\vb{k}}^m$ and the influence of higher order correlations neglected in the resonance triplet model will be the subject of future work. 

The molecular pairing field $\kappa_{\vb{k}}^m$ is four-body in nature, and thus negligibly small over the time regimes we consider. This is especially true for broad resonances, where its amplitude is further suppressed by the short molecular lifetime. It is included in the resonance triplet model for the sake of completeness at the level of second order correlations, and because its simulation is trivial in terms of the computational cost, yet allows us to confirm the above expectations numerically as well. We find that, for all analysis presented in the main text, $\kappa_{\vb{k}}^m$ can be safely excluded without altering the results.

\section{THREE-BODY EQUATIONS IN SHORT-RANGE OR VACUUM LIMITS}

In this section, we show that the equations of motion as introduced in Sec. \ref{sec:eom} converge to the three-body Schr\"odinger equation in the short-range or vacuum limits. The analogous convergence to the two-body Schr\"odinger equation in the resonance doublet model has been shown in Ref. \cite{Braun2022}, and may be compared to the similar convergence of the BCS gap equation \cite{Leggett2006}. Subsequently we discuss the numerical solution to the vacuum equations, and show that the binding energy of the Efimov trimer can be recognized as a non-universal scale in our many-body dynamics.

\subsection{Derivation of coupled-channel three-body Schrödinger equation}
\label{sec:schrod}
Consider a general three-body state $\ket*{\Psi}$ with energy $E$, such that $\hat{H} \ket*{\Psi} = E \ket*{\Psi}$. In the two-channel scenario, this state has four components,
\begin{align}
\ket*{\Psi} = \left(\ket*{\Psi_{aaa}}, \ket*{\Psi_{acc}}, \ket*{\Psi_{cac}}, \ket*{\Psi_{cca}} \right)^{T}.
\end{align}
Here $\ket*{\Psi_{aaa}}$ denotes the component with all three particles in the open channel and $\ket*{\Psi_{acc}}, \ket*{\Psi_{cac}}, \ket*{\Psi_{cca}}$ are the atom-dimer components which constitute the closed three-body channel. The open and closed channel states are coupled through the two-body interaction $V_{PQ}$ as introduced in Sec.~\ref{sec:renorm}, such that the three-body Schr\"odinger equation can be written as, 
\begin{align}
\begin{split}
E \begin{pmatrix}
\ket*{\Psi_{aaa}} \\ 
\ket*{\Psi_{acc}} \\ 
\ket*{\Psi_{cac}} \\ 
\ket*{\Psi_{cca}} \\ 
\end{pmatrix} = \begin{pmatrix}
\hat{H}_{aaa,aaa} & \hat{H}_{aaa,acc} & \hat{H}_{aaa,cac} & \hat{H}_{aaa,cca} \\
\hat{H}_{acc,aaa} & \hat{H}_{acc,acc} & 0 & 0 \\
\hat{H}_{cac,aaa} & 0 &  \hat{H}_{cac,cac} & 0 \\
\hat{H}_{cca,aaa} & 0  & 0 & \hat{H}_{cca,cca} \\
\end{pmatrix}.
\end{split}
\label{eq:SEq3B}
\end{align}
We now adopt a three-particle plane wave basis $\ket*{\vb{k},\vb{q},\vb{q}'}$, and calculate the projection $\Psi_{a}(\vb{k}, \vb{q}) = \braket*{\vb{k}, \vb{q}, -\vb{k} - \vb{q}}{\Psi_{aaa}}$. From equation \eqref{eq:SEq3B}, the closed-channel components of the three-body state can be directly related to $\ket*{\Psi_{aaa}}$ as,
\begin{align}
\begin{split}
\ket*{\Psi_{acc}} = \frac{1}{E - \hat{H}_{acc,acc}} H_{acc, aaa} \ket*{\Psi_{aaa}}.
\end{split}
\end{align}
We now insert a full set of eigenfunctions of $H_{acc,acc}$. Since the closed channel only contains the single molecular state $\ket*{\phi}$, we then obtain,
\begin{align}
\begin{split}
\ket*{\Psi_{acc}} &= \sum_{\vb{k}} \frac{\ket*{\vb{k}, \phi_{23}} \bra*{\vb{k}, \phi_{23}}}{E -  \varepsilon_{\vb{k}}^a -   \varepsilon_{\vb{k}}^m }H_{acc, aaa} \ket*{\Psi_{aaa}},
\end{split}
\end{align}
Here $\ket*{\vb{k}, \phi_{23}}$ defines a product state in which a molecular state $\ket*{\phi_{23}}$ composed of particles 2 and 3 moves as a plane wave with center-of-mass momentum $-\vb{k}$. Meanwhile, atom 1 moves similarly as a plane wave with momentum $\vb{k}$. We now define a momentum-space closed channel wave function $\Phi_{23}(\vb{k}) = \braket*{\vb{k}, \phi_{23}}{\Psi_{acc}}/\sqrt{2}$, and express the coupling potential in the separable form $H_{acc, aaa} = \frac{1}{\sqrt{V}}\ket*{\zeta_{23}} \beta \bra*{\zeta_{23}}$, consistent with the conversion of atoms 2 and 3 into a closed-channel molecule. Repeating the above procedure for the other closed channel components $\ket*{\Psi_{cac}}, \ket*{\Psi_{cca}}$ and projecting Eq.~\eqref{eq:SEq3B} on the open-channel subspace, we finally obtain, 
\begin{align}
\begin{split}
E \Psi_a(\vb{k} , \vb{q}) &= \left[\varepsilon_{\vb{k}}^a  + \varepsilon_{\vb{q}}^a  +  \varepsilon_{\vb{k} + \vb{q}}^a \right]\Psi_a(\vb{k} , \vb{q})  +  \frac{\sqrt{2}}{\sqrt{V}}\zeta_{23}(\vb{k} - \vb{q}) \beta \braket*{\zeta_{23}}{\phi_{23}} \Phi_{23}(\vb{k} + \vb{q}) \\ & + \frac{\sqrt{2} }{\sqrt{V}}\zeta_{31}(2\vb{q} + \vb{k}) \beta \braket*{\zeta_{31}}{\phi_{31}} \Phi_{31}(-\vb{k}) + \frac{\sqrt{2}}{\sqrt{V}} \zeta_{12}(-2\vb{k} - \vb{q}) \beta \braket*{\zeta_{12}}{\phi_{12}} \Phi_{12}(-\vb{q}).
\end{split}
\end{align}
Since we consider indistinguishable particles we have $\ket*{\zeta_{23}}  = \ket*{\zeta_{31}}  = \ket*{\zeta_{12}} $, $\ket*{\phi_{23}} = \ket*{\phi_{31}}  = \ket*{\phi_{12}}$ and $\Phi_{23} = \Phi_{31}  = \Phi_{12}$, such that we can suppress the particle indices going forward. We reintroduce the coupling constant $g = \sqrt{2} \beta \braket*{\phi}{\zeta}$, and obtain the coupled system,
\begin{align}
\begin{split}
E \Psi_a(\vb{k} , \vb{q}) &= \mathcal{S} \left\{\varepsilon_{\vb{k}}^a \Psi_a(\vb{k} , \vb{q}) + \frac{g}{\sqrt{V}} \zeta(\vb{k} - \vb{q}) \Phi(\vb{k} + \vb{q})  \right\}, \\
E \Phi(\vb{k}) &= \left(\varepsilon_{\vb{k}}^a  + \varepsilon_{\vb{k}}^m \right) \Phi(\vb{k}) + \frac{g}{2 \sqrt{V}} \sum_{\vb{q}} \zeta(2\vb{q} + \vb{k}) \Psi_a(\vb{q}, \vb{k} ).
\end{split}
\label{eq:CSet1}
\end{align}
Here we have used parity symmetry, i.e. $\Psi_{a}(-\vb{q}, -\vb{k}) = \Psi_{a}(\vb{q}, \vb{k})$. Equation \eqref{eq:CSet1} constitutes a set of two coupled (integral) equations that solve the three-body problem. We note that this set is formally equivalent to the equations derived in Ref. \cite{Gogolin2008}, although we obtain them by a different route.

\subsection{Equations of motion in short-range or vacuum limits}

We now consider the coupled evolution of the three-body cumulants $R_{\vb{k}, \vb{q}}^a$ and $\kappa_{\vb{k}, \vb{q}}^{am}$ in the limit of large momenta (short range). Then we find the equations of motion,
\begin{align}
\begin{split}
i \hbar \pdv{t}R_{\vb{k},\vb{q}}^a &= \mathcal{S}\bigg\{ \varepsilon_{\vb{k}}^a  R_{\vb{k},\vb{q}}^a + \frac{g}{\sqrt{V}} \zeta(\vb{k} - \vb{q}) \kappa_{\vb{k} + \vb{q}}^{am} \bigg\}, \\
i \hbar \pdv{t}\kappa_{\vb{k}}^{am} &= \left(\varepsilon_{\vb{k}}^a + \varepsilon_{\vb{k}}^m \right)\kappa_{\vb{k}}^{am} +  \frac{g}{2\sqrt{V}} \sum_{\vb{q}}  \zeta( 2\vb{q} + \vb{k}) R_{\vb{k},\vb{q}}^a.
\end{split}
\label{eq:RaKapam_sr}
\end{align}
The above set is fully equivalent to Eq. \eqref{eq:CSet1} upon adopting the ansatz,
\begin{align}
\begin{split}
R_{\vb{k}, \vb{q}}^a = e^{-i \frac{E}{\hbar} t} \Psi_a(\vb{k} , \vb{q}), \qquad \kappa_{\vb{k}}^{am} = e^{-i \frac{E}{\hbar} t} \Phi(\vb{k}).
\end{split}
\label{eq:QuasiStationary}
\end{align}
Hence the short-range limit of our cumulant model converges to the three-body Schr\"odinger equation as expected. Note that the set \eqref{eq:RaKapam_sr} can also be obtained by taking the vacuum limit where all populations vanish. At longer range or in the presence of a medium the ansatz in Eq. \eqref{eq:QuasiStationary} corresponds to an approximation where the density effects are treated as quasi-stationary \cite{Kira2011, Kira2015, Colussi2018_2}.

\subsection{Efimov trimer spectrum}

\begin{figure}[t]
\centering
\subfloat[\label{fig:ks_Rs}]{{\includegraphics[width=2.36in]{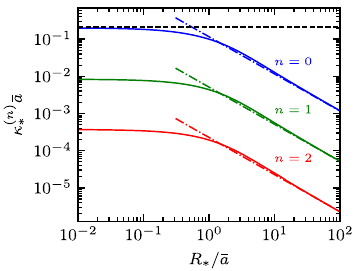}}}
\subfloat[\label{fig:C3fits}]{\includegraphics[width=2.36in]{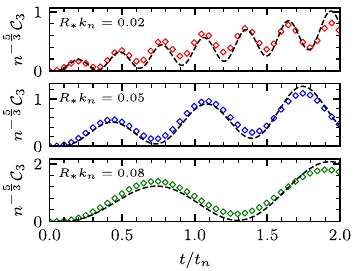}}
\subfloat[\label{fig:TrimerFreq}]{\includegraphics[width=2.36in]{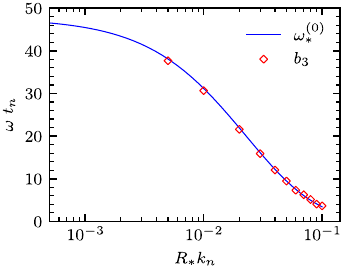}}
\caption{\label{fig:rhok_exp} Illustrations of Efimov physics embedded in the resonance triplet model. In (a) we plot the three-body binding wavenumber $\kappa_*^{(n)} = \sqrt{m\abs*{E_*^{(n)}}/\hbar^2}$ as a function of $R_*$ for the three lowest lying Efimov trimer states $n=0,1,2$, obtained from a numerical solution of Eq. \eqref{eq:STM}. The black dashed line shows the van der Waals universal three-body parameter $\kappa_*^{(0)} \bar{a} = 0.226$ for the ground state trimer obtained with a single-channel interaction, valid for $R_* \ll \bar{a}$. Dash-dotted lines show the theoretical prediction from Ref. \cite{Gogolin2008}, which holds in the narrow resonance limit $R_* \gg \bar{a}$. In (b) we show the Efimovian oscillation of the three-body contact density $\mathcal{C}_3$, which increases in frequency as the Feshbach resonance broadens in accordance with panel (a). To extract the exact frequency, we repeat the procedure in Ref. \cite{Colussi2020} and fit the growth of $\mathcal{C}_3$ to a function $f(t) = b_1 t^2 + b_2 \sqrt{t} \sin^2(b_3 t/2 - b_4)$. The obtained value of $b_3$ then represents the characteristic oscillation frequency, which we plot in panel (c) and compare directly with the ground state Efimov frequency $\omega_*^{(0)} = \abs*{E_*^{(0)}}/\hbar$ extracted from panel (a). The excellent match shows that the three-body Schr\"odinger equation is correctly included in the many-body dynamics.}
\end{figure}

In principal, one can now solve Eq. \eqref{eq:CSet1} to obtain the binding energies $E_*^{(n)}$ of the Efimov trimers at unitarity. However, since the two-body problem with our separable potential admits an analytic solution, it is advantageous to instead formulate the three-body equations in terms of the two-body transition matrix. First, we generalize Eq. \eqref{eq:tau_open} off the energy shell \cite{Secker2021_2},
\begin{align}
\begin{split}
\tau_{PP}(z) = \frac{\beta^2 \matrixel*{\zeta}{\hat{G}_0^{\mathrm{2B}}(z)}{\zeta}}{1 - \beta^2  \matrixel*{\zeta}{\hat{G}_0^{\mathrm{2B}}(z)}{\zeta} \matrixel*{\zeta}{\hat{G}_0^{\mathrm{2B}}(z- E_c)}{\zeta} - v_c \matrixel*{\zeta}{\hat{G}_0^{\mathrm{2B}}(z- E_c)}{\zeta}},
\end{split}
\end{align}
which we only have to evaluate for $z < 0$. Here $E_c$ is the threshold energy of the closed channel, and $v_c$ is the closed-channel interaction strength such that $\hat{V}_{QQ} = \frac{1}{\sqrt{V}} v_c \ket*{\zeta} \bra*{\zeta}$. Both $E_c$ and $v_c$ are fixed by choosing the asymptotic energetic separation between the channels at the bare resonance position $\nu = 0$. The Efimov spectrum is independent of this energy, provided that it is chosen large enough such that the closed-channel is far away from a shape resonance.

Then, in order to obtain the trimer binding energy $E_* < 0$, we follow the Faddeev approach and decompose the three-body state $\ket*{\Psi}$ as \cite{Faddeev1960, Glockle1983, Faddeev1993}, 
\begin{align}
\begin{split}
\ket*{\Psi} = \hat{\mathcal{S}}\ket*{\bar{\Psi}}.
\end{split}
\end{align}
The components $\ket*{\bar{\Psi}}$ are obtained from the Faddeev equations,
\begin{align}
\begin{split}
\ket*{\bar{\Psi}} = \hat{G}_0^{\mathrm{3B}}(E_*) \mathcal{T}(E_*)\hat{P} \ket*{\bar{\Psi}}
\label{eq:Faddeev}
\end{split}
\end{align}
which is an integral version of the three-body Schrödinger equation, with the boundary condition that the three-body wavefunction vanishes at infinity. Here $\hat{G}_0^{\mathrm{3B}}(E)$ is the uncoupled three-body Green's function, $\mathcal{T}(E)$ is the two-body transition matrix generalised to the three-body space, and $\hat{P} = \hat{P}_+ + \hat{P}_-$ \cite{Glockle1983}. Equation \eqref{eq:Faddeev} may be projected onto momentum states, after which it can be reduced to a one-dimensional integral equation,
\begin{align}
\begin{split}
\mathcal{F}(k) + \frac{1}{V} \sum_{\vb{q}} \frac{\zeta(\abs*{\vb{k} + 2\vb{q}}) \zeta(\abs*{2\vb{k} +\vb{q}})}{\kappa_*^2 + k^2 + q^2 + \vb{k} \cdot \vb{q}} \tau\left(-\frac{\hbar^2 \kappa_*^2}{m} - \frac{3}{4} \frac{\hbar^2 q^2}{m}\right) \mathcal{F}(q) = 0,
\end{split}
\label{eq:STM}
\end{align}
where $\kappa_* = \sqrt{m \abs{E_*}/\hbar^2}$. After transferring to the continuum limit the angular part of the integration can be done analytically, after which $\kappa_*$ is obtained numerically by solving the integral equation using Gaussian quadrature. The result is shown for a range of $R_*/\bar{a}$ in Fig. \ref{fig:ks_Rs}. In the narrow resonance limit $R_* \gg \bar{a}$, we also plot the analytical prediction $\kappa_* R_* = 2.6531$ mod $e^{\pi/s_0}$ obtained in Ref. \cite{Gogolin2008}. We additionally compare the result with the ground state three-body parameter $\kappa_* \bar{a} = 0.226$ obtained from a single-channel interaction, correspondent with the broad resonance limit $R_* \ll \bar{a}$. Both limits match well with our numerical results.

\section{GROWTH OF THE EXCITED STATE FRACTION}
\label{sec:excgrowth}

\begin{figure}[t]
\centering
\subfloat{{\includegraphics[width=\textwidth]{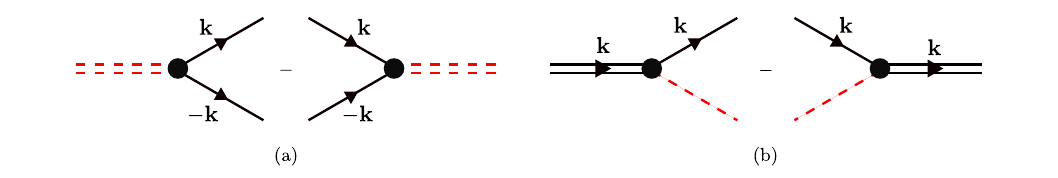}}}
\caption{\label{fig:diag_S12} Diagrammatic representation of $s_{\vb{k}}^{(1)}$ in (a) and $s_{\vb{k}}^{(2)}$ in (b). Atomic and molecular momenta shown with single and double lines respectively. Atomic(molecular) states shown with single(double) lines, and condensed states shown in dashed red. Since we take the imaginary part, each source consists of the balance between forward and backward processes.}
\end{figure}

From Eq.~\eqref{eq:nka} we recognize that the atomic excitation density is sourced by two distinct scattering processes. To make this explicit, we write $ \hbar \pdv{t} n_{\vb{k}}^a = s_{\vb{k}}^{(1)} + s_{\vb{k}}^{(2)}$, where,
\begin{align}
s_{\vb{k}}^{(1)} = 2 g \zeta(2 \vb{k}) \mathrm{Im} \left(\psi_m \kappa_{\vb{k}}^{a*} \right), \qquad
s_{\vb{k}}^{(2)} = -2 g \zeta(2 \vb{k}) \mathrm{Im} \left(\psi_a \chi_{\vb{k}} \right).
\end{align}
Diagrammatic representations of $s_{\vb{k}}^{(1/2)}$ are shown in Fig. \ref{fig:diag_S12}.
For now we focus on the first source $s_{\vb{k}}^{(1)}$. To examine the tail of the momentum distribution, we expand the formal solution of the equation of motion \eqref{eq:kappaa} for $\kappa_{\vb{k}}^a$ in powers of $1/k$. Then,
\begin{align}
s_{\vb{k}}^{(1)} &\underset{k \rightarrow \infty}{\longrightarrow} \frac{2 m^2 g^2}{ \hbar^4 k^4}  \mathrm{Im} \left(i \psi_m \pdv{t} \psi_m^*  \right)-\frac{4m g^2}{ \hbar^2 k^2} \mathrm{Im}(\psi_m \kappa_{\vb{k}}^{am*}  \psi_a). \label{eq:src_1_ana}
\end{align}
Evidently the tail of $s_{\vb{k}}^{(1)}$ is splits into a two-body and a three-body contribution. The diagrammatic representations of the associated processes are shown in Fig.~2 of the main text. As we show in Sec.~\ref{sec:contacts}, the two-body contribution can be directly related to the two-body contact density $\mathcal{C}_2$.

\begin{figure}[t]
\centering
\subfloat[\label{fig:MomRes_S1}]{{\includegraphics[width=3.4in]{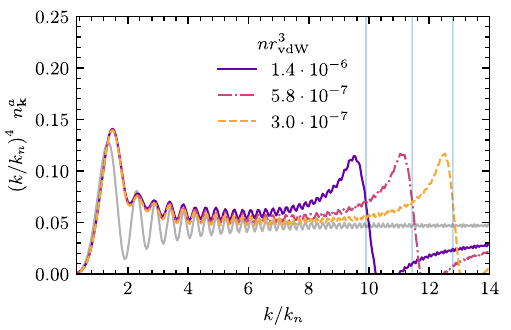}}}
\subfloat[\label{fig:MomRes_S2}]{\includegraphics[width=3.4in]{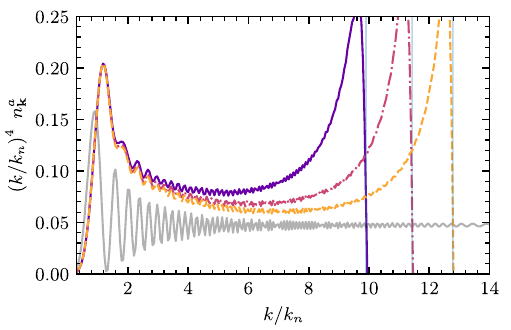}}
\caption{\label{fig:MomRes} Single-particle momentum distribution $n_{\vb{k}}^a$ for different values of the density scale $n r_{\mathrm{vdW}}^3$, at $t = t_n$ in (a) and $t = 2 \ t_n$ in (b). Computed with resonance width $R_* k_n = 0.4$. The universal doublet curve is shown in grey, and the positions $k_{\mathrm{res}}$ of the resonances due to $s_{\vb{k}}^{(2)}$ are shown with vertical lines. As the density of the gas decreases the resonance moves to higher momenta, signifying its finite-range nature. At later times the width of the resonance also increases since it scales with the magnitude of $\kappa_{\vb{k}}^{am}$.}
\end{figure}

We now consider the second source $s_{\vb{k}}^{(2)}$, which is neglected in the resonance doublet model. Its diagrammatic representation is shown in Fig. \ref{fig:diag_S12}(b). From the equation of motion \eqref{eq:chi} for $\chi_{\vb{k}}$ we derive that $s_{\vb{k}}^{(2)}$ is largest for momenta $k$ close to $k_{\mathrm{res}} = \sqrt{4m\nu/\hbar^2}$, where the process in Fig. \ref{fig:diag_S12}(b) is resonant, meaning that the change in kinetic energy is exactly equal to the change in internal energy. Far away from $k_{\mathrm{res}}$, we find,
\begin{align}
s_{\vb{k}}^{(2)} &\underset{\abs{k - k_{\mathrm{res}}} \rightarrow \infty}{\longrightarrow}   -\frac{8 m g^2}{\hbar^2 (k_{\mathrm{res}}^2 - k^2)}\mathrm{Im}\left( \psi_a \kappa_{\vb{k}}^{am*} \psi_m \right).
\label{eq:src_2_ana}
\end{align}
Since $k_{\mathrm{res}} \sim \sqrt{\Lambda}$, $s_{\vb{k}}^{(2)}$ will vanish in the formal zero-range limit $\Lambda \rightarrow \infty$. For finite values of $\Lambda$, $s_{\vb{k}}^{(2)}$ induces an asymmetric ``resonance" feature in the tail of the momentum distribution, see Fig.~\ref{fig:MomRes}. As we also mentioned in Sec.~\ref{sec:eom}, this resonance compensates for the additional energy introduced into the system via $\kappa_{\vb{k}}^{am}$, ensuring that the total energy is always conserved. Evidently however, the actual shape of the peak has unphysical features, most notably a negative excitation density for momenta $k > k_{\mathrm{res}}$. As $n_{\vb{k}}^m$ is sourced purely by $s_{\vb{k}}^{(2)}$, its value is similarly unphysical in the current model. The negativity can be attributed to the truncation of the cumulant hierarchy, in which the open channel analogue of $\chi$, $\expval*{a^{\dagger}a^{\dagger}a}$, is neglected. Inclusion of this cumulant however, induces hierarchical couplings to higher order correlations that significantly complicate the model. For the scope of the present work, it is sufficient that the non-universal effects of $s_{\vb{k}}^{(2)}$ are negligible for the momenta of interest $k/k_n<4$ which is indeed the case as shown in Fig. \ref{fig:MomRes} and indicates that our results in Fig.~4 of the main text are universal. In addition we note that the population densities $n_{k_{\mathrm{res}}}$ form a practically negligible fraction of the total population ($\sim 10^{-5} \ n$).

\section{CONTACT RELATIONS IN A RESONANCE MODEL}
\label{sec:contacts}

The two- and three-body contact densities $\mathcal{C}_2$ and $\mathcal{C}_3$ can be directly related to short-range correlation functions using an Effective Field Theory (EFT) \cite{Braaten2006, Braaten2011}. Here the single-channels two-body interaction is replaced by a local coupling constant $g_2$, which can be directly related to the scattering length,
\begin{align}
a = \frac{m g_2}{8 \pi \hbar^2} \left(1 + \frac{ m g_2 \Lambda}{4 \pi \hbar^2}  \right)^{-1}.
\end{align}
From Eq.~\eqref{eq:aRs} we observe that the low-energy two-body scattering properties of the resonance model are equivalent with EFT following the mapping,
\begin{align}
g_2 = - \frac{m}{\hbar^2}\frac{g^2}{\nu}.
\label{eq:g2map}
\end{align}
By inserting this expression for $g_2$ into the contact relations quoted in Ref.~\cite{Braaten2008} we obtain Eqs. (8) and (9) in the main text. The validity of the mapping Eq.~\eqref{eq:g2map} can also be argued from a more fundamental starting point by adiabatically eliminating the molecular condensate field $\psi_m$ from the dynamics, which reproduces the single-channel HFB equations with the coupling strength given by Eq.~\eqref{eq:g2map} \cite{Braun2022}. At equilibrium adiabatic elimination is justified provided that $R_* k_n (\bar{a} k_n)^2 \ll 1$, which is always true in our calculations \cite{Gurarie2007}. We have confirmed numerically that it is also valid out-of-equilibrium following the quench, except at very early times $t \ll t_*$ where the molecular condensate is small. In a similar way adiabatic elimination of $\kappa_{\vb{k}}^{am}$ in Eq.~\eqref{eq:RaKapam_sr} gives the vacuum three-body Schr\"odinger equation, again with two-body coupling strength given by Eq.~\eqref{eq:g2map}.

To obtain $\mathcal{C}_2$ and $\mathcal{C}_3$ we expand $\expval{d^{\dagger} d}$ and $\expval{t^{\dagger} t}$ into cumulants. We define the Fourier transformations at zero particle separation as,
\begin{align}
\begin{split}
n_a(\vb{0}) = \frac{1}{V} \sum_{\vb{k}} n_{\vb{k}}^a, \quad \kappa_a(\vb{0}) = \frac{1}{V} \sum_{\vb{k}} \kappa_{\vb{k}}^a,  \quad R_a(\vb{0},\vb{0}) = \frac{1}{V^{\frac{3}{2}}} \sum_{\vb{k},\vb{q}} R_{\vb{k},\vb{q}}^a
\end{split}
\end{align}
Then we find,
\begin{align}
\begin{split}
\expval{d^{\dagger} d} &= \abs*{\psi_a}^4 + 4 \abs{\psi_a}^2 n_a(\vb{0}) + 2 n_a^2(\vb{0}) + \abs{\kappa_a(\vb{0})}^2 + \mathrm{Re} \left[ \psi_a^2 \kappa_a^*(\vb{0}) \right]
\end{split}
\end{align}
and,
\begin{align}
\begin{split}
\expval{t^{\dagger} t} &=  \abs{\psi_a}^6 + 9 \abs{\psi_a}^4 n_a(\vb{0}) + 6 \abs{\psi_a}^2 \mathrm{Re}\left[ \psi_a^{*2} \kappa_a(\vb{0}) \right] + 2 \mathrm{Re} \left[\psi_a^{*3} R_a(\vb{0}, \vb{0}) \right]  + 9 \abs*{\psi_a}^2 \abs{\kappa_a(\vb{0})}^2 \\& \qquad + 18 \abs*{\psi_a}^2 n_a^2(\vb{0}) + 18 \mathrm{Re}\left[\psi_a^2 \kappa_a^*(\vb{0}) n_a(\vb{0}) \right] + 6 \mathrm{Re}\left[\psi_a \kappa_a(\vb{0}) R_a^{*}(\vb{0}, \vb{0})\right] + \abs{R_a(\vb{0}, \vb{0})}^2 \\& \qquad + 9 \abs{\kappa_a(\vb{0})}^2 n_a(\vb{0}) + 6 n_a^3(\vb{0}).
\end{split}
\end{align}
In the zero-range limit $\Lambda \rightarrow \infty$, the scaling of the different cumulants dictates that \cite{Colussi2020},
\begin{align}
\begin{split}
\expval{d^{\dagger} d} \underset{\Lambda \rightarrow \infty}{\longrightarrow} \abs{\kappa_a(\vb{0})}^2, \qquad \expval{t^{\dagger} t} \underset{\Lambda \rightarrow \infty}{\longrightarrow} \abs{R_a(\vb{0}, \vb{0})}^2,
\end{split} 
\end{align}
giving Eq. (12) in the main text. This behavior in the zero-range limit is also confirmed by our numerics. 

It is instructive to note that Eq. (9) in the main text can be derived analytically for the resonance doublet model, starting from the asymptotic behavior of the source $s_{\vb{k}}^{(1)}$ as given in Eq. \eqref{eq:src_1_ana}, and adiabatically eliminating the molecular field. Then we can rewrite Eq. \eqref{eq:src_1_ana} as,
\begin{align}
\begin{split}
s_{\vb{k}}^{(1)} &\underset{k \rightarrow \infty}{\longrightarrow} \frac{2 m^2 g^2}{ \hbar^4 k^4}  \mathrm{Im} \left(i  \psi_m \pdv{t} \psi_m^*  \right) = \frac{m^2 g^4}{ \hbar^4 k^4}  \pdv{t} \abs{\psi_m}^2 = \frac{m^2 g^4}{4 \hbar^4 \nu^2} \frac{1}{k^4}  \pdv{t} \abs{\frac{1}{V} \sum_{\vb{q}} \kappa_{\vb{q}}^a}^2
\end{split}
\label{eq:src_analytic}
\end{align}
Upon integrating this expression over time one obtains $n_{\vb{k}}^a \rightarrow \mathcal{C}_2/k^4$, where $\mathcal{C}_2$ is the two-body contact density. Hence Eq. (9) in the main text is contained within the doublet cumulant equations, consistent with the convergence seen in the upper panels of Fig. 4 in the main text.

\section{CONDENSATE FRACTIONS}

In Ref.~\cite{MusolinoPRL} it was shown that the contact densities following the quench can be directly related to macroscopic order parameters that signify condensation of two-body and three-body composites. To show that the growth of $\mathcal{C}_3$ in our model is indeed associated with the development of a significant triple condensate, we repeat this analysis here for the resonance model. First we separate out single-particle condensation by decomposing the field operators as $\hat{\psi}_{\sigma}(\vb{r}) = \sqrt{V} \psi_{\sigma} + \delta \hat{\psi}_{a}(\vb{r})$, with $\expval*{\delta \hat{\psi}_{\sigma}(\vb{r})} = 0$ and $\sigma = a,c$ to denote the open and closed channels respectively. Note that since particles in the closed channel only appear in pairs we always have $\psi_c = 0$. We define condensation of a $p$-body composite by evaluating the existence of off-diagonal long-range ordering (ODLRO) in the $p$-body density matrix, given by \cite{Braun2022},
\begin{align}
\rho^{(p)}( \mathbf{r}_1' \sigma_1', \ldots \mathbf{r}_{p}' \sigma_{p}',\mathbf{r}_1 \sigma_1, \ldots \mathbf{r}_{p} \sigma_p;t) = \expval*{ \prod_{i=1}^p \delta \psi_{\sigma_i'}^{\dagger}(\mathbf{r}_i') \prod_{i=1}^p \delta \psi_{\sigma_i'}(\mathbf{r}_i) }.
\label{eq:rho_pbody}
\end{align}
Here $\sigma = a,c$ for open and closed channel operators, and we recall that the number of closed channel operators must be even. There exists a spectral decomposition \cite{Yang1962},
\begin{align}
\rho^{(p)}( \mathbf{r}_1' \sigma_1', \ldots \mathbf{r}_{p}' \sigma_{p}',\mathbf{r}_1 \sigma_1, \ldots \mathbf{r}_{p} \sigma_p;t) = \sum_{\nu} N^{(p)}_{\nu}(t) \varphi_{\nu}^{(p)  *}(\mathbf{r}_1' \sigma_1', \ldots \mathbf{r}_{p}' \sigma_{p}',t)  \varphi_{\nu}^{(p)}(\mathbf{r}_1 \sigma_1, \ldots \mathbf{r}_{p} \sigma_p,t),
\label{eq:rho_sd}
\end{align}
where  $\varphi_{\nu}^{(p)}$ represent orthonormal eigenvectors of the density matrix with eigenvalues $N_{\nu}^{(p)}$. As before, we project the closed-channel onto the single resonant bound state $\ket*{\phi}$, which removes all states with unpaired closed channel particles. The existence of $p$-body ODLRO is associated with a single nonzero eigenvalue $N_{0}^{(p)}$ in the long-range (LR) limit $\abs{\sum_{i=1}^p \vb{r}_i - \vb{r}_i'}/p \rightarrow \infty$. In this regime the correlator on the right hand side of Eq. \eqref{eq:rho_pbody} is dominated by the anomalous contractions $\expval*{\delta \psi^{\dagger} \ldots \delta \psi^{\dagger}} \expval*{\delta \psi \ldots \delta \psi}$ \cite{Leggett2006}. Specifically one finds at the two-body level \cite{Braun2022},
\begin{align}
\rho^{(2)}(\mathbf{r}'\sigma_1' \sigma_2',\mathbf{r} \sigma_1 \sigma_2;t) \underset{\mathrm{LR}}{\longrightarrow} N^{(2)}_{0}(t)  \varphi_{0}^{(2)*}(\mathbf{r}'\sigma_1' \sigma_2',t)  \varphi_{0}^{(2)}(\mathbf{r} \sigma_1 \sigma_2,t),
\end{align}
where,
\begin{align}
N_0^{(2)}(t) = \sum_{\vb{k}} \abs{\kappa_{\vb{k}}^a}^2 + 2 \abs{\psi_m}^2, \qquad \underline{\varphi}_0^{(2)}(\vb{r}, t)  = \frac{1}{\sqrt{N_0^{(2)}(t)}} \begin{bmatrix} \kappa_a(\vb{r}) \\
\sqrt{2} \phi(\vb{r}) \psi_m \end{bmatrix}, 
\end{align}
with $\vb{r} = \vb{r}_1 - \vb{r}_2$, $\phi(\vb{r}) = \frac{1}{\sqrt{V}} \sum_{\vb{k}} e^{i \vb{k} \cdot \vb{r}} \phi(\vb{k})$, and $\kappa_a(\vb{r}) = \frac{1}{V} \sum_{\vb{k}} e^{i \vb{k} \cdot \vb{r}} \kappa_{\vb{k}}^a$. For notational convenience we write $\underline{\varphi}_0^{(2)}(\vb{r}, t)$ as the vector of spin components $\varphi_{0}^{(2)}(\mathbf{r} \sigma_1 \sigma_2,t)$, where $\sigma_1\sigma_2 = aa$ and $\sigma_1\sigma_2 = cc$ for the open and closed channel respectively. On the three-body level we have the allowed spin states $\sigma_1 \sigma_2 \sigma_3 = \left\{aaa, acc, cac, cca \right\}$, c.f. Sec. \ref{sec:schrod}. In the long-range limit, we obtain,
\begin{align}
\rho^{(3)}(\mathbf{r}'\boldsymbol{\rho}'\sigma_1' \sigma_2' \sigma_3',\mathbf{r}\boldsymbol{\rho} \sigma_1 \sigma_2 \sigma_3;t) \underset{\mathrm{LR}}{\longrightarrow} N^{(3)}_{0}(t) \varphi_{0}^{(3)*}(\mathbf{r}'\boldsymbol{\rho}'\sigma_1' \sigma_2' \sigma_3',t)  \varphi_{0}^{(3)}(\mathbf{r} \boldsymbol{\rho} \sigma_1 \sigma_2 \sigma_3,t),
\end{align}
where,
\begin{align}
\label{eq:triplewf}
N_0^{(3)}(t) = \sum_{\mathbf{k},\mathbf{q}} \abs{R^a_{\mathbf{k},\mathbf{q}}}^2 + 6 \sum_{\mathbf{k}} \abs{\kappa^{am}_{\mathbf{k}}}^2, \qquad \underline{\varphi}_0^{(3)}(\mathbf{r},\boldsymbol{\rho},t) = \frac{1}{\sqrt{N_0^{(3)}}} \begin{bmatrix} R_a(\vb{r}, \vb*{\rho}) \\ \sqrt{2} \phi(\vb{r}) \kappa_{am}(\vb*{\rho})  \\ \sqrt{2}  \phi\left(\vb{r}_{+}\right) \kappa_{am}(\vb*{\rho}_+) \\ \sqrt{2} \phi\left(\vb{r}_-\right) \kappa_{am}(\vb*{\rho}_-)  \end{bmatrix}.
\end{align}
Here we use the atom-dimer separation $\vb*{\rho} = \vb{r}_3 - (\vb{r}_1 + \vb{r}_2)/2$ such that $(\vb{r}, \vb*{\rho})$ together constitute a set of Jacobi coordinates. The sets $(\vb{r}_{\pm}, \vb*{\rho_{\pm}})$ are obtained after a (anti-)cyclic permutation of the particle indices. The Fourier transformed cumulants read $R_a(\vb{r}, \vb*{\rho}) = \frac{1}{V^{3/2}}\sum_{\mathbf{k},\mathbf{q}} e^{i \left(\frac{1}{2} \vb{k} + \vb{q}\right) \cdot \vb{r}} e^{i \vb{k} \vb*{\rho}} R^a_{\mathbf{k},\vb{q}}$ and $\kappa_{am}(\vb*{\rho}) = \frac{1}{V} \sum_{\mathbf{k}} e^{i \cdot \mathbf{k} \cdot \boldsymbol{\rho}} \kappa^{am}_{\mathbf{k}}$.

\begin{figure}[t]
\centering
\includegraphics[width=3.4in]{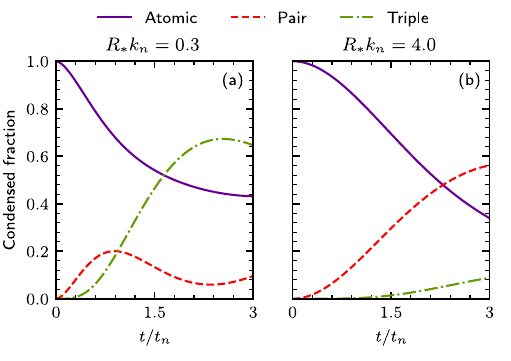}
\caption{\label{fig:CondFrac} Dynamics of the single, pair and triple condensate fractions following the quench, shown for a broad Feshbach resonance in (a) and a narrow Feshbach resonance in (b).}
\end{figure}

Contrary to the atomic condensate, the eigenvalues $N_0^{(p)}$ can not be directly interpreted as condensed fractions due to Bose enhancement inside the few-body complex \cite{MusolinoPRL}. Mathematically, this discrepancy appears as a violation of the canonical commutation relations for the bosonic composite operators,
\begin{align}
\hat{d}_0^{(p)}= \frac{1}{\sqrt{p!}} \sum_{\sigma_1 \hdots \sigma_p} \left[\prod_{i=1}^p \int d^3\vb{r}_i \delta \hat{\psi}_{\sigma_i}(\vb{r}_i) \right] \varphi_0^{(p) }(\vb{r}_1 \sigma_1, \ldots, \vb{r}_p \sigma_p),
\end{align}
which annihilate few-body composites. From Eq. \eqref{eq:rho_sd} it follows that $\expval*{\hat{d}_0^{(p) \dagger} \hat{d}_0^{(p)}} = N_0^{(p)}/p!$. Assuming a nonzero fraction of excited atoms, we find the following commutation relations,
\begin{align}
\left\langle \left[\hat{d}_0^{(2)},\hat{d}_0^{(2) \dagger} \right] \right\rangle &= 1 + \frac{2}{N_0^{(2)}} \sum_{\vb{k}} \abs{\kappa_{\vb{k}}^a}^2 n_{\vb{k}}^a, \\
\left\langle \left[\hat{d}_0^{(3)},\hat{d}_0^{(3) \dagger} \right] \right\rangle &= 1+\frac{1}{N_0^{(3)}} \sum_{\mathbf{k},\mathbf{q}} 3 n^a_{\mathbf{k}}(1+n^a_{\mathbf{q}})\abs{R^a_{\mathbf{k},\mathbf{q}}}^2 + \frac{6}{N_0^{(3)}} \sum_{\mathbf{k}} (n^a_{\mathbf{k}}+n^m_{\mathbf{k}})  \abs{\kappa^{am}_{\mathbf{k}}}^2.
\end{align}
Following Ref. \cite{MusolinoPRL}, we define renormalized operators $\hat{D}_0^{(p)} = \hat{d}_0^{(p)}/\sqrt{\expval*{[\hat{d}_0^{(p)},\hat{d}_0^{(p) \dagger} ]}}$ and define the pair and triple condensate fractions of the total density as $n_0^{(p)}/n = \expval*{\hat{D}_0^{(p) \dagger} \hat{D}_0^{(p)}}/(N/p)$. By this method we ensure that the condensed composites are on average bosonic, thus avoiding overcounting due to Bose enhancement. The dynamics of the single, pair and triple condensate fractions are shown in Fig. \ref{fig:CondFrac}, for the resonance strengths also plotted in Fig.~3 of the main text.
As expected we observe a significant fraction of condensed triples for the broad resonance, while the number of condensed pairs is reduced. This is consistent with the behavior of $\mathcal{C}_2$ and $\mathcal{C}_3$ characterized in the main text. However, one will note that even with the rebosonisation of the pair and triple operators the condensed fractions are still unphysically large, with a total $>1$. This discrepancy originates from the negative population of excited molecules that appears due to the unphysical source $s_{\vb{k}}^{(2)}$, discussed in Sec.~\ref{sec:excgrowth}. Because of this negative population, the number of excited atoms at these large momenta is correspondingly larger than it should be, such that there will be an unphysical number of high-momenta excitations that can go into the pair and triple condensates. Still, given the rapid decrease of population with momentum, the dynamics in Fig. \ref{fig:CondFrac} should be predominantly seeded by the low momentum modes, where the model is physical. Hence we expect that a higher-order model which restores the positivity of the excitation densities will give similar qualitative results to Fig.~\ref{fig:CondFrac}.

\bibliography{References_sup}